\documentclass[aps,amsmath,showpacs,prl,twocolumn]{revtex4}

\usepackage{graphicx}
\usepackage{amssymb}
\newcommand{\ba}{\begin{eqnarray}}
\newcommand{\ea}{\end{eqnarray}}
\newcommand{\bmath}{\begin{mathletters}}
\newcommand{\emath}{\end{mathletters}}
\newcommand{\ban}{\begin{eqnarray*}}
\newcommand{\ean}{\end{eqnarray*}}

\newcommand{\bsub}{\begin{subequations}}
\newcommand{\esub}{\end{subequations}}

\begin{document}

\title{Partial Dynamical Symmetry at Critical-Points of 
Quantum Phase Transitions}

\author{A. Leviatan}

\affiliation{
Racah Institute of Physics, The Hebrew University,
Jerusalem 91904, Israel}

\date{\today}

\begin{abstract}
We show that partial dynamical symmetries (PDS) 
can occur at critical-points 
of quantum phase transitions, in which case, 
underlying competing symmetries are conserved exactly 
by a subset of states, 
and mix strongly in other states. 
Several types of PDS are demonstrated with the example of
critical-point Hamiltonians for first- and second-order transitions 
in the framework of the interacting boson model, 
whose dynamical symmetries correspond to different shape-phases 
in nuclei. 
\end{abstract}


\maketitle

Symmetries play a profound role in determining the critical 
behaviour of dynamical systems. Their significance 
was recognized in Landau's classic 
theory of thermal phase 
transitions~\cite{Landau} and in the renormalization group 
of critical phenomena~\cite{RG}. 
An equally-important role is played by symmetries in quantum phase 
transitions (QPT) or 
ground-state energy phase transitions~\cite{gilmore79}, 
which occur at zero temperature as a function of a 
coupling constant. Such structural changes are currently 
of great interest 
in different branches of physics. 
QPT occur as a result of a competition between terms in the Hamiltonian  
with different symmetries which lead to 
considerable mixing in the eigenfunctions, especially 
at the critical-point where the structure changes most rapidly. 
In the present work we address the question whether there are any 
symmetries (or traces of) still present at the critical-points of QPT. 
We show that particular 
symmetry constructions, called partial dynamical symmetries,  
can survive at the critical-point in-spite of the strong mixing. 
The feasibility of such persisting symmetries 
gains support from the recently proposed~\cite{iac0001} and empirically  
confirmed~\cite{caszam0001} analytic descriptions of critical-point nuclei, 
and the emergence of ``quasi-dynamical symmetries''~\cite{rowe0405} 
in the vicinity of such critical-points.

A convenient framework to study symmetry-aspects of QPT are models 
where the Hamiltonian is expanded in elements 
of a Lie algebra ($G_0$), called the spectrum generating algebra. 
A dynamical symmetry occurs if the Hamiltonian
can be written in terms of the Casimir operators 
of a chain of nested algebras of $G_0$, 
\ba
G_0\supset G_1\supset \ldots \supset G_n ~, 
\label{ds}
\ea
terminating with an invariant algbera $G_n$. 
The following properties are then observed. 
(i)~All states are solvable and analytic expressions
are available for energies and other observables. 
(ii)~All states are classified by quantum numbers, 
$\vert\alpha_0,\alpha_1,\ldots,\alpha_n\rangle$, 
which are the labels of the irreducible representations (irreps)
of the algebras in the chain. 
(iii)~The structure of wave functions is completely dictated by symmetry
and is independent of the Hamiltonian's parameters. 
Partial dynamical symmetry (PDS) corresponds to
a particular symmetry-breaking for which some (but not all) of the above
mentioned virtues of a dynamical symmetry are retained. 
PDS of type~I corresponds to a situation where 
{\it some} of the states have {\it all} the dynamical symmetry. 
In this case the properties (i)-(iii) 
are fulfilled exactly, but 
by only a subset of states. 
PDS of type~II corresponds to a situation for which 
{\it all} the states preserve {\it part} of the dynamical symmetry. 
In this case there are no analytic solutions,
yet selected quantum numbers (of the conserved symmetries) are retained.
This can occur, for example, 
when, the Hamiltonian preserves only 
selected symmetries $G_i\subset G_n$ in the 
chain~(\ref{ds}), and only their irreps are unmixed. 
PDS of type~III has a hybrid character, 
for which {\it some} of the 
states preserve {\it part} of the dynamical symmetry. 
PDS of various types 
have been shown to be relevant to nuclear and molecular 
spectroscopy~\cite{lev96,isa99,levvan02,
esclev00,rowe01,pinchen97} and to 
mixed systems with coexisting regularity and chaos~\cite{whelan93}. 
All examples of PDS encountered so far involved stable limits of 
structure. In the present work we show 
the relevance of the PDS notion 
to the more complicated case of a phase transition. 

As a concrete example, we consider the interacting boson model 
(IBM)~\cite{ibm}, widely used 
in the description of quadrupole collective states in 
nuclei, in terms of a system of $N$ monopole ($s$) and
quadrupole ($d$) bosons, representing valence nucleon pairs. 
The spectrum generating algebra is $G_0=U(6)$ 
and the invariant algebra is 
$G_{n}=O(3)$. 
The three dynamical symmetry limits 
of the model and corresponding bases are
\begin{subequations}
\ba
U(6) &\supset& U(5)  \supset O(5) \supset O(3)
\qquad 
\vert N,n_d,\tau,\tilde{\nu},L\rangle \qquad\label{u5ds} \\
U(6) &\supset& SU(3) \supset O(3)
\qquad\qquad\;\;\,
\vert N,(\lambda,\mu),K,L\rangle \qquad\label{su3ds} \\
U(6) &\supset& O(6)  \supset O(5) \supset O(3)
\qquad
\vert N,\sigma,\tau,\tilde{\nu},L\rangle~. \qquad
\label{o6ds}
\ea
\end{subequations}
\begin{figure}[t]  
\begin{center}
\rotatebox{270}{\includegraphics[scale=0.35]
{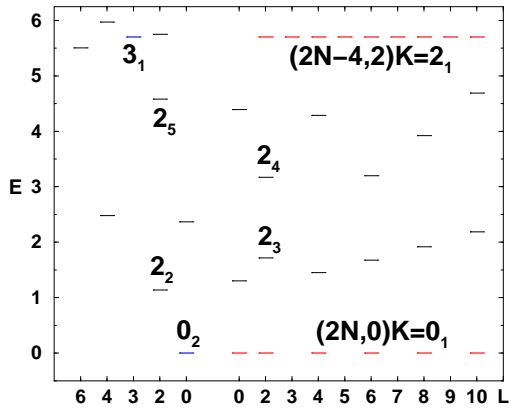}}\hspace{1cm}
\end{center}
\vspace{-0.2cm}
\caption{(color online). 
Spectrum of $H(\beta_0=\sqrt{2})$, Eq.~(\ref{hcri1st}), 
with $h_2=0.1$ and $N=10$. 
$L=0_2,3_1$ are solvable $U(5)$ states of Eq.~(\ref{spher}). 
$L(K=0_1)$ and $L(K=2_1)$ are, respectively, solvable $SU(3)$ states 
of Eq.~(10a) and Eq.~(10b) with $k=1$.
\label{fig1}}
\end{figure}
The quantum numbers $N,n_d,(\lambda,\mu),\sigma,\tau$ and $L$ 
label the relevant irreps of $U(6),U(5),SU(3),O(6),O(5)$ and $O(3)$ 
respectively. 
$\tilde{\nu}$ and $K$ are multiplicity-labels needed for complete 
classification of selected states in the reductions $O(5)\supset O(3)$ and 
$SU(3)\supset O(3)$ respectively. 
The analytic solutions 
of these dynamical symmetries 
resemble a spherical vibrator, axially-deformed rotor and 
deformed $\gamma$-soft rotor for the $U(5)$, $SU(3)$ and $O(6)$ chains 
respectively.
This identification is consistent with the geometric visualization 
of the model in terms of a potential surface defined by 
the expectation value of the Hamiltonian in the coherent (intrinsic) 
state~\cite{gino80,diep80}
\ba
\vert\beta,\gamma ; N \rangle &=&
(N!)^{-1/2}(b^{\dagger}_{c})^N\,\vert 0\,\rangle ~,
\label{cond}
\ea
where 
$b^{\dagger}_{c} = (1+\beta^2)^{-1/2}[\beta\cos\gamma 
d^{\dagger}_{0} + \beta\sin{\gamma} 
( d^{\dagger}_{2} + d^{\dagger}_{-2})/\sqrt{2} + s^{\dagger}]$. 
For the general 
IBM Hamiltonian with one- and two-body interactions, the 
potential surface reads 
\ba
E(\beta,\gamma) = E_0 + 
N(N-1)
\frac{\left [ a\beta^{2} - b\beta^3\cos 3\gamma + c\beta^4\right ]}
{(1+\beta^2)^2} ~.
\label{eint}
\ea
The coefficients $E_0,a,b,c$ involve particular linear 
combinations of the Hamiltonian's parameters~\cite{lev87}. 
The quadrupole shape parameters $(\beta,\gamma)$ at the global minimum
of $E(\beta,\gamma)$ define the equilibrium shape for a given Hamiltonian. 
The shape can be spherical $(\beta=0)$ or deformed $(\beta>0)$ 
with $\gamma=0$ (prolate), $\gamma=\pi/3$ (oblate), 
or $\gamma$-independent $(b=0)$. 

Phase transitions can be studied 
by IBM Hamiltonians of the form, $H_1 + gH_2$, involving 
terms from different dynamical symmetry chains~\cite{diep80}. 
The nature of the phase transition is 
governed by the topology of the corresponding surface (\ref{eint}), 
which serves as a Landau's potential 
with the equilibrium deformations as order parameters. 
The conditions 
on the surface at the critical-points of first- and second-order 
transitions are 
\bsub
\ba
&&1^{st}\, {\rm order}\qquad\;\;\;
b^{2}=4ac,\;a>0,\; b\neq 0\qquad\qquad\qquad\\
&&2^{nd}\, {\rm order}\qquad\;\;\;
\, a=0,\; b=0,\; c>0 ~.
\ea
\label{1st2nd}
\esub
\begin{figure}[t]
\begin{center}
\rotatebox{270}{\includegraphics[scale=0.4]
{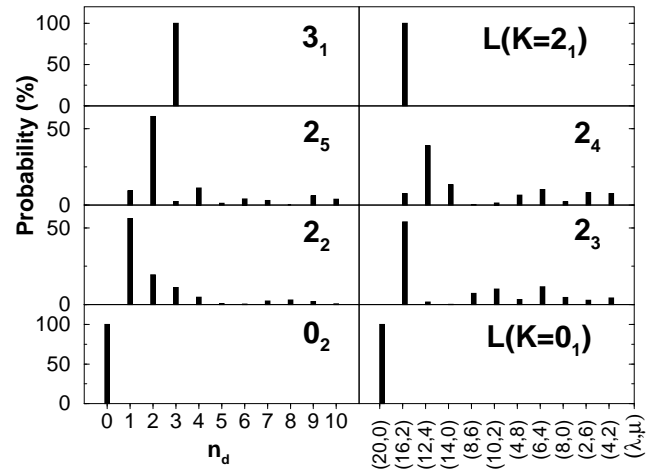}}
\end{center}
\vspace{-0.5cm}
\caption{$U(5)$ ($n_d$) and $SU(3)$ $[(\lambda,\mu)]$ decomposition of 
selected spherical and deformed states in Fig.~1.
\label{fig2}}
\end{figure}
The first-order critical-surface~(5a) has degenerate spherical and 
deformed minima at $\beta=0$ and ($\beta=2a/|b|,\gamma_0$), 
with $\gamma_0=0\,(\pi/3)$ for $b>0\,(b<0)$. 
The second-order critical-surface~(5b) is independent of $\gamma$ 
and behaves as $\beta^4$ for small $\beta$. 
The conditions in Eq.~(5) fix the critical value of the control 
parameter $(g=g_c)$ which, in turn, determines the critical-point 
Hamiltonian. IBM Hamiltonians of this type have been 
used extensively for studying shape-phase transitions in 
nuclei~\cite{iaczam04,rowe0405,levgin03,lev05,lev06}. 
We now show that a large class of such Hamiltonians 
exhibit PDS.

The spherical to deformed $\gamma$-soft shape-phase transition 
is modeled in the IBM by
the Hamiltonian
\ba
H &=& \epsilon\,\hat{n}_d + 
A \left[\, d^{\dagger}\cdot d^{\dagger} -  (s^{\dagger})^2\,\right ]
\left[\, H.c.\,\right]
\nonumber\\
\epsilon &=& 4(N-1)A ~,
\label{hcri2nd}
\ea
where $H.c.$ stands for Hermitian conjugate and
the dot implies a scalar product. 
The $\hat{n}_d$-term is the
$d$-boson number operator (eigenvalues $n_d$), which is the linear 
Casimir operator of $U(5)$. The $A$-term is related to the Casimir 
operator of $O(6)$~\cite{ibm}. 
For the indicated ratio of coefficients,  
the above $H$ satisfies 
condition~(5b), hence 
qualifies as a second-order critical Hamiltonian. 
The first (second) term in Eq.~(\ref{hcri2nd}) 
has $O(6)$ [$U(5)$] selection rules $\Delta\sigma=0,\pm 2$ 
($\Delta n_d=0,\pm 2$), 
and both terms are $O(5)$-scalars. 
Consequently, the eigenstates of $H$ 
have good $O(5)$ symmetry $(\tau)$, but 
are mixed strongly with respect to both $U(5)$ and 
$O(6)$~\cite{levgin03}. Since both $U(5)$ and $O(6)$ are broken 
while $O(5)\supset O(3)$ are preserved, 
by definition, the critical Hamiltonian has
an $O(5)$ PDS of type II. In fact, since $O(5)$ is a good symmetry 
common to both chains (2a) and (2c), the $O(5)$ PDS is valid 
throughout the $U(5)$-$O(6)$ transition region. 
\begin{figure}[b]
\begin{center}
\rotatebox{270}{\includegraphics[scale=0.35]
{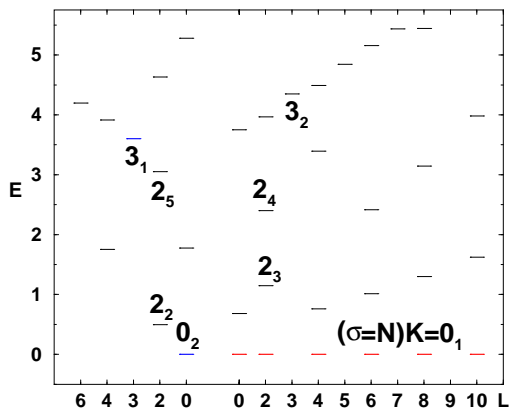}}\hspace{1cm}
\end{center}
\vspace{-0.2cm}
\caption{(color online). 
Spectrum of $H(\beta_0=1)$, Eq.~(\ref{hcri1st}), 
with $h_2=0.1$ and $N=10$. 
$L=0_2,3_1$ are solvable $U(5)$ states of Eq.~(\ref{spher}). 
$L(K=0_1)$ are solvable states of Eq.~(\ref{solvo6}) with good 
$O(6)$ but broken $O(5)$ symmetry.
\label{fig3}}
\end{figure}

A recent study of QPT within the IBM has shown that, apart from 
rotational terms which do not affect the potential surface 
of Eq.~(\ref{eint}), the critical Hamiltonian for a spherical 
to prolate-deformed shape-phase transition can be transcribed 
in the form~\cite{lev06}
\ba
H(\beta_0) &=& h_{2}\, 
P^{\dagger}_{2}(\beta_0)\cdot\tilde{P}_{2}(\beta_0) ~,
\label{hcri1st}
\ea 
where
$P^{\dagger}_{2\mu}(\beta_0) = 
\beta_{0}\,s^{\dagger}d^{\dagger}_{\mu} + 
\sqrt{7/2}\,\left( d^{\dagger} d^{\dagger}\right )^{(2)}_{\mu}$, 
$\tilde{P}_{2\mu}(\beta_0)=(-1)^{\mu}P_{2,-\mu}(\beta_0)$ 
and $h_2,\,\beta_0>0$. 
The corresponding surface in Eq.~(\ref{eint}) has coefficients  
$a=h_2\beta_{0}^2, b=2h_{2}\beta_{0},c=h_2$, 
which satisfy condition~(5a). This qualifies $H(\beta_0)$ 
as a first-order critical Hamiltonian whose potential accommodates 
two degenerate minima at $\beta=0$ and $(\beta,\gamma)=(\beta_0,0)$. 
$H(\beta_0)$ is constructed to have the equilibrium intrinsic state,
$\vert\beta=\beta_0,\gamma=0 ; N\rangle$, Eq.~(\ref{cond}), 
as a zero-energy eigenstate. 
Rotational-invariance 
ensures that states, $\vert\beta_0;N,L\rangle$, 
of good $O(3)$ symmetry $L$ projected from this intrinsic state, 
remain zero-energy eigenstates. $H(\beta_0)$ then has 
a solvable deformed ground band, 
\ba
\vert\beta_0;N,L\rangle \quad E=0\qquad\;\; (L=0,2,4,\ldots, 2N)~.
\qquad\quad
\label{deform}
\ea
It has also the following solvable spherical eigenstates 
\bsub
\ba 
\vert N,n_d=\tau=L=0 \rangle  \;\; &&E = 0\\
\vert N,n_d=\tau=L=3 \rangle \;\;
&&E = 3 h_2[\beta_{0}^2 (N-3) + 5]~.
\qquad\quad
\ea
\label{spher}
\esub
As shown in Figs.~1-4, 
the remaining states in the spectrum of 
$H(\beta_0)$ are either predominantly 
spherical (with characteristic dominance of single $n_d$ 
components) or deformed states (with a broad $n_d$ distribution) 
arranged in several excited bands. 
\begin{figure}[b]
\begin{center}
\rotatebox{270}{\includegraphics[scale=0.4]
{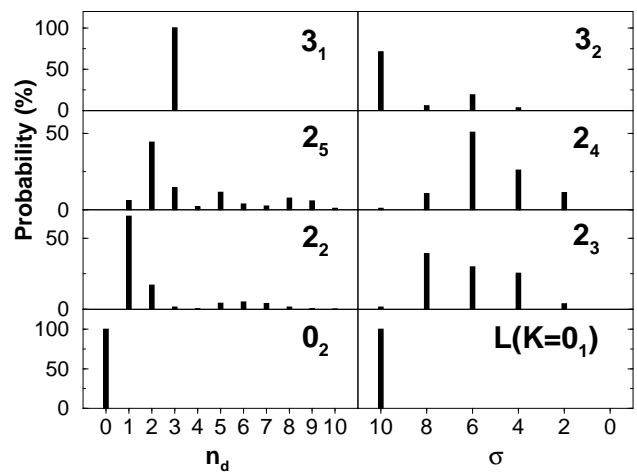}}\hspace{1cm}
\end{center}
\vspace{-0.5cm}
\caption{$U(5)$ ($n_d$) and $O(6)$ $(\sigma)$ decomposition of 
selected spherical and deformed states in Fig.~3.
\label{fig4}}
\end{figure}

The critical Hamiltonian of Eq.~(\ref{hcri1st}) with $\beta_0=\sqrt{2}$ 
is a special case of a Hamiltonian shown in~\cite{lev96} to have 
$SU(3)$ PDS. 
This comes about because the sequence of states 
$\vert k\rangle \propto [P^{\dagger}_{2,2}(\sqrt{2})]^{k}
\vert \beta=\sqrt{2},\gamma=0;N-2k\rangle$
are eigenstates of $H(\beta_0=\sqrt{2})$. 
These are lowest-weight states 
in the $SU(3)$ irreps $(\lambda,\mu)=(2N-4k,2k)$ with $2k\leq N$. 
In the nuclear physics terminology they are referred to as intrinsic 
states representing deformed ground $(k=0)$ and $\gamma^k$ 
bands, with angular momentum projection $(K=2k)$ along the symmetry axis. 
Since $H(\beta_0=\sqrt{2})$ is an $O(3)$-scalar, the states of 
good $L$ projected from 
$\vert k\rangle$ remain eigenstates with 
quantum numbers, $\vert N,(\lambda,\mu)K,L\rangle$, 
of the $SU(3)$ chain~(2b), and form solvable bands, 
\bsub
\ba
\vert N,(2N,0)K=0,L\rangle 
\;\; E=0
\;\; (L=0,2,4,\ldots, 2N)\quad\,\,\;\;\\
\vert N,(2N-4k,2k)K=2k,L\rangle
\;\; 
E=3h_2[2N+1-2k]k\;
\nonumber\\
L=K,K+1,K+2,\ldots, (2N-2k)\qquad k>0~.\quad\quad
\label{solvsu3}
\ea
\esub
In addition, $H(\beta_0=\sqrt{2})$ has the spherical states of 
Eq.~(\ref{spher}), with good $U(5)$ symmetry, as eigenstates. 
The remaining levels of $H(\beta_0=\sqrt{2})$, shown 
in Fig.~1, are calculated numerically. 
Their wave functions are spread over many 
$U(5)$ and $SU(3)$ irreps, as is evident from Fig.~2. This situation, 
where some states are solvable with good $U(5)$ symmetry, 
some are solvable with good $SU(3)$ symmetry and all other 
states are mixed with respect to both $U(5)$ and $SU(3)$, 
defines a $U(5)$ PDS of type I 
coexisting with a $SU(3)$ PDS of type I.

The Hamiltonian of Eq.~(\ref{hcri1st}) with $\beta_0=1$ is a special 
case of a Hamiltonian shown in~\cite{levvan02} to have $O(6)$ PDS. 
This comes about because the intrinsic state of Eq.~(\ref{cond}) with 
$(\beta_0=1,\gamma=0)$ is a zero-energy eigenstate of $H(\beta_0=1)$ 
with  good $O(6)$ symmetry $(\sigma=N)$. The $O(3)$-invariance 
of the Hamiltonian ensures that states of good $L$ projected from 
$\vert\beta_0=1,\gamma=0;N\rangle$ 
form a solvable ground band with good $O(6)$ character,
\ba
\vert N,\sigma=N,L\rangle \qquad E=0
\qquad (L=0,2,4,\ldots, 2N)~.
\quad
\label{solvo6}
\ea
In addition, $H(\beta_0=1)$ has the spherical states of 
Eq.~(\ref{spher}), with good $U(5)$ symmetry, as eigenstates. 
The remaining eigenstates in Fig.~3 are mixed with respect to both 
$U(5)$ and $O(6)$, as is evident from their 
decomposition shown in Fig.~4. Apart from the solvable $U(5)$ states 
of Eq.~(\ref{spher}), all eigenstates of $H(\beta_0=1)$ are mixed 
with respect to $O(5)$ (including the solvable 
$O(6)$ states of Eq.~(\ref{solvo6}), as shown in Fig.~5). 
It follows that the Hamiltonian has a subset of states 
with good $U(5)$ symmetry and a subset of states with good $O(6)$ 
but broken $O(5)$ symmetry, and all other states are mixed with respect 
to both $U(5)$ and $O(6)$. These are precisely the required features of 
$U(5)$ PDS of type I coexisting with $O(6)$ PDS of type III.
\begin{figure}[t]
\begin{center}
\vspace{-0.2cm}
\rotatebox{270}{\includegraphics[scale=0.3]
{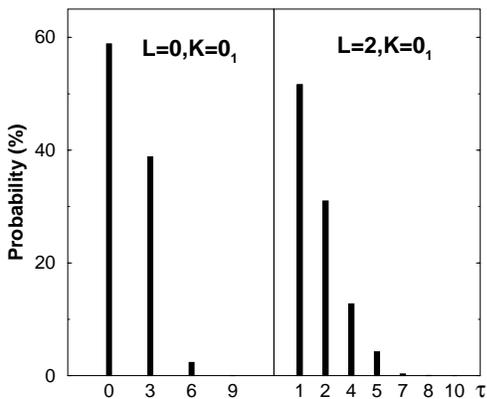}}\hspace{1cm}
\end{center}
\vspace{-0.5cm}
\caption{$O(5)$ ($\tau$) decomposition of the $L=0,\,2$ states, 
Eq.~(\ref{solvo6}), members 
of the ground band ($K=0_1$) shown in Fig.~3.
Both states have $O(6)$ symmetry $\sigma=N$.
\label{fig5}}
\end{figure}

For arbitrary values of $\beta_0$,
the spherical states with good $U(5)$ symmetry, Eq.~(\ref{spher}), 
are still eigenstates of $H(\beta_0)$, Eq.~(\ref{hcri1st}), 
hence $U(5)$ PDS of type I is still valid. 
In general, the deformed states of Eq.~(\ref{deform}), 
are not associated with any IBM dynamical symmetry 
but, nevertheless, are still solvable. This situation may be referred to 
as partial solvability. 
Since the wave functions of the solvable states are 
known, it is possible to obtain closed form expressions for 
related observables. For example, for the electromagnetic 
$E2$ operator, 
$T(E2)=d^{\dagger}s+s^{\dagger}\tilde{d} 
+ \chi(d^{\dagger}\tilde{d})^{(2)}$, 
the necessary matrix elements for transitions involving 
states in Eqs.~(\ref{deform})-(\ref{spher}) are 
$T_1\equiv\langle \beta; N, L \vert\vert\, T(E2)\, 
\vert \vert \beta; N,L+2\rangle$ and
$T_2\equiv\langle \beta; N, L=2\vert\vert\, T(E2)\, 
\vert\vert N,n_d=\tau=L=0\rangle$, 
\ba
T_1 &=& 
C_L\,\frac{\beta
[\,a_1\,\Gamma_{N-1}^{(L)}(\beta) 
+ a_2\, \Gamma_{N-1}^{(L+2)}(\beta)\,]}
{[\Gamma_{N}^{(L)}(\beta)\,\Gamma_{N}^{(L+2)}(\beta)]^{1/2}}~, 
\nonumber\\
T_2 &=& 
\beta N/
[ N!\,\Gamma^{(2)}_{N}(\beta)\,]^{1/2}~,
\label{t1t2}
\ea
where  
$C_L=\sqrt{2L+5}\,(L+2,0;2,0\vert L,0)$ is proportional to a 
Clebsch Gordan coefficient, $a_1= 1-\beta\bar{\chi}L/(2L+3)$, 
$a_2= 1-\beta\bar{\chi}(L+3)/(2L+3)$ 
with $\bar{\chi}=\sqrt{2/7}\chi$, 
and $\Gamma_{N}^{(L)}(\beta)$ is a normalization factor given 
in~\cite{lev06}.
 
As discussed, the spectrum of $H(\beta_0)$, Eq.~(\ref{hcri1st}), exhibits 
coexistence of spherical and deformed states, signaling a first-order 
transition. In particular, the spherical $L=0$ state, Eq.~(9a), is 
exactly degenerate with the ground band, Eq.~(\ref{deform}), 
and for $\beta_0=\sqrt{2}$ also the spherical $L=3$ state, Eq.~(9b), 
is degenerate with the $SU(3)$ $\gamma$-band, Eq.~(10b) with $k=1$. 
Adding to the Hamiltonian the Casimir operator of $O(3)$, contributes 
an exact $L(L+1)$ splitting with no effect on wave functions. 
The remaining degeneracy of states with the same $L$, can be lifted 
by adding a small one-body term $\hat{n}_d$. 
With that, the spherical $U(5)$ states of Eq.~(\ref{spher}) 
remain solvable eigenstates. However, the $\hat{n}_d$ term destroys 
the exact solvability and partial-symmetry of the deformed states, 
Eq.~(\ref{deform}). 
The corresponding leading-order shifts can be estimated from 
$\langle \beta; N,L\vert \hat{n}_d
\vert \beta; N,L \rangle=
N - \Gamma_{N-1}^{(L)}(\beta)/\Gamma_{N}^{(L)}(\beta)$. 

In summary, we have shown the relevance of the PDS notion 
to critical-points of QPT, with phases characterized by Lie-algebraic 
symmetries. In the example considered, second-order critical Hamiltonians 
mix incompatible symmetries 
but preserve a common lower symmetry, resulting in a single PDS 
with selected quantum numbers conserved. 
First-order critical Hamiltonians 
exhibit distinct subsets of solvable states with good symmetries, 
giving rise to a coexistence of different PDS. 
The ingredients of an algebraic description 
of QPT is a spectrum generating algebra and an associated geometric 
space, formulated in terms of coherent (intrinsic) states. 
The same ingredients are used in the construction of Hamiltonians 
with PDS. These, in accord with the present work, can be used as tools 
to explore the role of possibly partial symmetries in governing 
the critical behaviour of diverse dynamical systems undergoing QPT. 
This work was supported by the Israel Science Foundation.

\end{document}